\documentclass[12pt,a4paper]{article}
\usepackage{graphicx}
\usepackage{epsfig}
\usepackage{amssymb}
\usepackage{amsmath}
 \textwidth 16true cm
\begin{document}

\begin{center}
\Large\textbf{Early Re-brightenings in GRB Afterglows As Signatures of
Low-to-High Density Boundary}\\
\vspace{0.5cm}

\large{P. H. Tam$^{a,}$\footnote{Corresponding author -- e-mail:
acthomas@graduate.hku.hk\\ {\it \,\,\,This work has been accepted to
be published in New Astronomy} }, C. S. J. Pun$^{a}$, Y. F.
Huang$^{b}$, K.
S. Cheng$^{a}$}\\ \ \\

\small\it{$^{a}$Department of Physics, The University of Hong Kong,
Pokfulam Road, Hong Kong, China.} \\
\small\it{$^{b}$Department of Astronomy, Nanjing University, Nanjing
210093, China.}
\end{center}

\begin{center}
\bf{ABSTRACT}
\end{center}
The association of long gamma-ray bursts (GRBs) with star forming
regions and the idea of massive stars as progenitors of GRBs are
widely accepted. Because of their short lifetimes, it is very likely
that massive stars are still embedded in dense molecular clouds when
they give birth to GRBs. Stellar winds from GRB progenitors can
create low-density bubbles with sizes and densities strongly
depending on the initial ambient density. A boundary between the
bubble and the dense molecular cloud must exist with the density at
the boundary increasing from that of the bubble to that of the outer
cloud. We have calculated the lightcurves of the afterglows in such
environments with three regions: the stellar wind region, the
boundary, and the molecular cloud. We show that the interaction
between the cylindrical jet and the density boundary can result in a
re-brightening of the afterglow occurring as early as $\sim 1$ day
after the GRB. We compare our models with the optical afterglows of
GRB~970508, GRB~000301C, and GRB~030226. We find that the values of
our model parameters, including the radius of the wind bubble, the
densities in the bubble and in the
outer molecular cloud are within typical ranges.  \\ \ \\
\section{Introduction}

The association of supernovae (SNe) with gamma-ray bursts (GRBs), of
which a convincing example is GRB~030329 -- SN~2003dh, gives strong
supports for massive stars as progenitors of GRBs (Hjorth et al.,
2003). It has been suggested that GRBs are located in star-forming
regions (Holland \& Hjorth, 1999; Holland et al., 2001; Lamb \&
Reichart, 2000) and most of the star formation in the universe
occurs in molecular clouds. Several observational evidences of the
presence of molecular clouds around GRB progenitors have been found.
Galama \& Wijers (2001) analyzed a sample of eight GRB afterglows.
They found X-ray evidence for column densities $N_{\rm
H}=10^{22}-10^{23}$~cm$^{-2}$ of gas around these GRBs. From dust
destruction and survival of neutral Mg, they further constrained the
clouds to be of masses $\gtrsim 10^{5}M_{\odot}$ and sizes within 10
to 30 parsecs. These values are consistent with measurements of
Galactic giant molecular clouds (GMCs; Solomon et al. 1987).
Moreover, Reichart \& Yost (2001) showed that dark bursts likely
occur in clouds of similar sizes. Reichart \& Price (2002) further
suggested that the bursts with detected optical afterglows also
occur in GMCs and have shown that the results are consistent with
observations statistically.

Observations reveal that GMCs ($M\sim10^{5}-10^{6}M_{\odot}$,
$R\sim6-60$~pc, $n_{\rm H}\sim10^{2}-10^{3}$~cm$^{-3}$) are highly
inhomogeneous and contain a large number of clumps or dense cores
($M\lesssim 10^{3}M_{\odot}$, $R\sim0.1-1$~pc, $n_{\rm
H}\sim10^{4}-10^{5}$~cm$^{-3}$). These dense clumps or cores are
believed to be the birth places of massive stars. In fact, such high
density environments have been used to explain the afterglows of
GRB~980519 and GRB~990123 (Dai \& Lu, 2000; Wang et al., 2000).

Jets in GRBs have been proposed to avoid large implied isotropic
energy in some GRBs and to explain the rapid decay in some afterglow
lightcurves (Rhoads, 1997, 1999; Sari et al., 1999; Huang et al.,
2000b, 2000c; Frail et al., 2001). However, almost all these
discussions have assumed a conical geometry (i.e. keeping the
half-opening angles of the jets largely constant, or allowing some
lateral expansion), which we think may not be sufficient.
Observations of some relativistic jets in radio galaxies, active
galaxies, and ``microquasars'' show that they are very likely to be
cylindrical (i.e. at large scales the cross-sectional areas of the
jets are constant or at most expand laterally with sound speed;
Perley et al., 1984; Biretta et al., 1999). Therefore it is
necessary to explore the behaviors of cylindrical jets in addition
to conical jets. Cheng et al. (2001) noted its significance and gave
analytic as well as numerical calculations of the cylindrical jet
model. Huang et al. (2002) showed that the afterglows of GRBs
970228, 970508, 971214, 980329, and 980703 could be fit by
cylindrical jets propagating into uniform medium. Cylindrical jets
propagating in free wind environments (i.e. $n\propto r^{-2}$) have
also been analyzed (Ma et al., 2003).

The physics of stellar winds around the GRB progenitors has widely
been considered in the literature (Chevalier \& Li, 1999, 2000;
Ramirez-Ruiz et al., 2001; Wu et al., 2003). The strong wind blown
out from a very massive star will dramatically affect the density
distribution in the circumstellar medium. In a dense cloud
($n_0\sim10^5$~cm$^{-3}$), the wind is able to drive a low density
bubble, which evolves to a radius of about $1$~pc in $\approx10^5$
years (Shull, 1980), beyond which situates a denser shell mainly
consisting of swept-up materials. Dai \& Lu (2002) analyzed the
hydrodynamics and emission features of GRB ejecta on a preburst
environment with a sudden density jump.

The increasing number of GRB afterglow lightcurves observed enables
us to model the fine details of their evolutions. One important
behavior is the re-brightenings (or ``bumps'') in the optical
afterglow lightcurves. They show excess flux compared to
extrapolation from earlier data, deviating from the predictions of
simple models which assume isotropic fireballs or collimated jets
and homogeneous interstellar medium (ISM) or simple wind environment
(i.e. $n\propto r^{-2}$). Many late-time bumps (e.g. GRBs~970228,
980326, and 011121) have been attributed to SN components because of
the reddening of the observed spectra (Bloom, 2003; Bloom et al.,
1999, 2002; Galama et al., 2000; Garnavich et al., 2003b). These
late-time bumps are also argued to be due to dust echoes (Esin \&
Blandford, 2000). On the other hand, GRB~970508 and GRB~000301C show
early-time optical bump(s) with no significant color change in the
spectra. A number of proposals to explain the early-bumps have been
raised. For example, energy injections and gravitational
microlensing have been suggested to explain the bump(s) in
afterglows of GRB~970508 (e.g. Panaitescu et al., 1998; Chang et
al., 2002) and GRB~000301C (Garnavich et al., 2000) respectively. In
this paper, we present a unified model to explain these two GRBs: a
cylindrical jet, after pushing through the low density bubble, can
produce the observed lightcurve bump when it encounters the
molecular cloud. GRB~030226 is another burst which reveals a
re-brightening during $t\sim0.2-0.5$~day (Dai \& Wu, 2003), thus can
also be interpreted by our model.

The structure of this paper is as follows. We outline the models in
section 2, emphasizing the three regions existed at the time of
GRBs, i.e. the wind bubble region, the intervening region, and the
molecular cloud. In section 3 we present numerical results showing
the effects of the density boundaries on the optical lightcurves,
which are characterized by observed bumps. We proceed in section 4
to compare our model with the optical afterglow lightcurves of GRBs
970508, 000301C, and 030226. Finally, we discuss our results and
present our conclusion in section 5.

\section{The Model}

\subsection{Density profile}

The strong wind blown out from a very massive star (OB-type main
sequence) will drastically affect the density distribution in the
circumstellar medium because of the large mass-loss rate and the
fast wind (Castor et al., 1975; Weaver et al., 1977). They
demonstrated that the stellar wind will create a low-density bubble.
When the mass of swept-up interstellar gas is comparable to the wind
mass, materials will accumulate at a certain radius, at where
situates a high density shell. The shell will then expand gradually
during the remaining life-time of the star. They showed that after
$t\approx10^6$~years, the bubble size can be about $20-30$~parsecs.
The bubble is much smaller (about $0.1-1.0$~pc) when the star is
embedded in a dense cloud ($n_{0}\sim10^{5}$~cm$^{-3}$; Shull,
1980). In this case, unshocked molecular gas remains outside the
shell with a particle density typical of that within a dense cloud
($n\sim10^3-10^5$~cm$^{-3}$). Ramirez-Ruiz et al. (2001) presented a
detailed analysis of stellar winds from single Wolf-Royet stars.
Ramirez-Ruiz et al. (2001) also considered the afterglows of GRBs
resulting from the interactions between Wolf-Royet winds and jets of
GRBs. They assumed that the jets of GRB have conical geometry and
found that re-brightenings can occur in afterglows about $10-100$
days after the GRBs. They suggested that this effect could explain
the observed bumps in afterglows of GRB~970228, GRB~980326, and
GRB~000911. However, the model cannot explain re-brightenings
occurring in less than 10 days. For example, GRB970508 has a bump in
the optical afterglow at $\sim 1$ day after the main burst.

We consider a simple model in this paper just to illustrate the
effects of density boundaries on the afterglow lightcurves. In this
model, the density $n_{1}$ is assumed to be constant for $R<R_{\rm
rise}$, where $R_{\rm rise}$ represents the onset of the boundary
between the stellar wind and the molecular cloud. We call this
Region (1). This region corresponds to the region of nearly constant
density in the simulated density profiles in Weaver et al. (1977)
and Ramirez-Ruiz et al. (2001). We assume $R_{\rm rise}$ to be
larger than the deceleration radius $R_{0}$, where the afterglow
phase starts. The density increases rapidly for a distance $d$ to a
much higher value $n_{2}$. We designate the density rising zone
Region (2) and the dense outer zone Region (3). Figure~1 illustrates
the different regions in our density model of the GRB surroundings.

\subsection{Kinetic Equations of Cylindrical Jet}

The idea of cylindrical jets has been supported by many observations
in systems other than GRBs. For example, jets in many radio galaxies
are cylindrical and maintain constant cross sections on large
scales. In addition, jets in many Herbig-Haro (HH) objects are
cylindrical (e.g. Ray et al., 1996). Observations show that HH jets
are poorly focused at first, but are collimated into cylinders at
sufficiently late times.

Theoretically, black hole-accretion disk systems can naturally
produce cylindrical jets (Shu et al., 1995; Krasnopolsky et al.,
2003; Vlahakis \& K\"{o}nigl, 2003a, 2003b; Fendt \& Ouyed, 2004),
with magnetic forces playing important roles in the collimation
process. The poloidal component of a dipolar magnetic field varies
as $B_P\propto r^{-3}$, where $r$ is the distance from the center.
Fendt \& Ouyed (2004) showed that the motion of matter along the
poloidal magnetic field lines will unavoidably produce a strong
toroidal field component, which decays as $B_{T}\propto r^{-1}$. As
a result, a magnetohydrodynamic (MHD) jet is dominated by the
toroidal component $B_T$ at large length scales. This field then
exerts an inward force on the MHD jet through ``hoop stress" and
provides the collimation. Many numerical results have shown that MHD
jets are initially conical. After the acceleration process, their
half opening angles become smaller and finally the jets become
cylindrical ones (e.g. Krasnopolsky et al., 2003). In the case of
GRBs, believed to occur in star formation regions, large density
gradients in the circumburst regions may also play a role in
collimation of the jets.

We therefore believe that the cylindrical jet model is worth
exploring in the case of GRBs, especially when one of the most
popular models now, the collapsar model (Woosley, 1993;
Paczy\'{n}ski, 1998), suggests the black hole-accretion disk system
as the progenitors of GRBs. Before calculating the afterglow
emissions, we need a unified dynamical model to describe the
evolution of GRB jets. In the internal-external shock model (Piran,
1999; Cheng \& Lu, 2001), afterglow is generated from the external
shock at which the ejecta of GRB is slowed down due to their
interactions with the surrounding materials. Such deceleration can
be described by the following equation, which are valid for both the
relativistic and non-relativistic phases (Huang et al., 1999a,
1999b):
\begin{equation}
\frac{d\gamma}{dm} = - \frac{\gamma^{2}-1}{M_{\rm ej}+\varepsilon m
+2(1-\varepsilon)\gamma m},
\end{equation}
where $\gamma$ is the Lorentz factor of the macroscopic motion of
the shocked material, $m$ the rest mass of the swept-up ISM, $M_{\rm
ej}$ the ejected mass from the progenitor, and $\varepsilon$ the
radiation efficiency. Here we assume an adiabatic jet and adopt
$\varepsilon=0$ throughout this paper (this is a good approximation
as long as $\xi\ll 1$). The evolution of the distance from the
progenitor $(R)$, the swept-up mass $(m)$ and the lateral radius
$(a)$ of the cylindrical jet can be described by the following
equations (Cheng et al., 2001):
\begin{equation}
\frac{dm}{dR} = \pi a^{2}nm_{p},
\end{equation}
\begin{equation}
\frac{dR}{dt} = \frac{\beta c}{1-\beta\cos\Theta},
\end{equation}
\begin{equation}
\frac{da}{dt} = \frac{v_{\bot}}{\gamma(1-\beta\cos\Theta)}
\end{equation}
where $n$ is the number density of hydrogen atoms with mass $m_{p}$
in the circumburst medium, $t$ the observer's time, $\beta =
\sqrt{1-1/\gamma^{2}}$, $\Theta$ is the angle between jet axis and
our line of sight, and $v_{\perp}$ the lateral expansion speed of
the jet in the comoving frame respectively. Since observations have
shown that the jets in other astrophysical objects are
well-collimated (Perley et al., 1984; Biretta et al., 1999; also see
Cheng et al., 2001), in this paper we will adopt $v_{\perp}=0$ (i.e.
the cylindrical jets strictly maintain a constant cross-sectional
area).

\subsection{Synchrotron Radiation}

The dominant radiation mechanism of the afterglows is synchrotron
radiation from the accelerated shocked material. The observed flux
density at frequency $\nu$ is given by (Huang et al., 2000a, 2000c)
\begin{equation}
S_{\nu} =  \frac{1}{\gamma^{3}(1-\beta cos\Theta)^{3}}
\frac{1}{4\pi D^{2}_{L}}P^{\prime}[\gamma(1-\beta cos\Theta)\nu],
\end{equation}
where $D_{L}$ is the luminosity distance, and
$P^{\prime}(\nu^{\prime})$ is the synchrotron power at
$\nu^{\prime}$ in the local frame given by Rybicki \& Lightman
(1979):
\begin{equation}
P^{\prime}(\nu^{\prime}) =
\frac{\sqrt{3}e^{3}B^{\prime}}{m_{e}c^{2}}
\int_{\gamma_{e,min}}^{\gamma_{e,max}}
\left(\frac{dN_{e}^{\prime}}{d\gamma_{e}}\right)F(\nu^{\prime}/\nu_{cr}^{\prime})d\gamma_{e},
\end{equation}
in which $B^{\prime}$ is the local magnetic field strength, $e$ the
electron charge, $\gamma_{e,max}=10^{8}/\sqrt{B^{\prime} ({\rm G})
}$,
$\gamma_{e,min}=\xi_{e}(\gamma-1)(m_{p}/m_{e})\frac{p-2}{p-1}+1$,
$F(x)=x\int_{x}^{\infty}K_{5/3}(k)dk$, where $K_{5/3}$ is a Bessel
function, and $\nu^{\prime}_{cr}=3\gamma_{e}^{2}eB^{\prime}/(4\pi
m_{e}c)$. The electron distribution follows a segmented power law,
where
\begin{equation}
\frac{dN_{e}^{\prime}}{d\gamma_{e}}\propto \begin{cases}
                    \gamma_{e}^{-p} &\text{if} \  \gamma_{e}<\gamma_{c}, \\
                    \gamma_{e}^{-(p+1)} &\text{if}\  \gamma_{e}\geq \gamma_{c}.
                  \end{cases}
\end{equation}
Here, $\gamma_{c}=6\pi m_{e}c/(\sigma_{T}\gamma B^{\prime 2}t)$ is
called the cooling Lorentz factor, meaning that an electron with
$\gamma_{e}<\gamma_{c}$ cools slowly and that with
$\gamma_{e}\geq\gamma_{c}$ cools rapidly and $\sigma_T$ is Thomson's
cross section. As usual, we assume that the magnetic energy density
is a fraction $\xi_{B}^{2}$ of the energy density, i.e. $B^{\prime
2}/8\pi=\xi_{B}^{2}e^{\prime}$, and that the electrons carry a
fraction $\xi_{e}$ of the energy. A more detailed description of the
electron distribution can be found in Dai et al. (1999) and Huang \&
Cheng (2003).

\section{Numerical Results}

In this section, we calculate the evolution of GRB jets to
understand the effects of a density rising region on the observed
lightcurves, especially how the parameters of this density rising
region (such as $n_2$, $R_{\rm rise}$, and $d$) affect the optical
afterglows. Progenitor models predict an ejected mass $M_{\rm
ej}\sim10^{-7}-10^{-9}M_{\odot}$, here we take $M_{\rm
ej}=10^{-8}M_{\odot}$. For the initial Lorentz factor, the electron
power-law index and the equipartition factors, we assume
conventional values: $\gamma_0 = 300$, $p = 2.2$, $\xi_e = 0.1$, and
$\xi_B^2 = 10^{-4}$. The cross-sectional radius is taken as $a_0 =
10^{14}$ cm as a reference value, as Cheng et al. (2001) do. Unless
otherwise specified, the parameters illustrated in Figure~1 are set
as $n_1 = 10$~cm$^{-3}$, $n_2 = 200$~cm$^{-3}$, $R_{\rm rise} =
3$~pc, and $d = 1$~pc. We place ourselves at $z=1$ from the source
and right on the jet axis of the cylindrical jet (an on-axis
observer with $\Theta = 0$). In this paper we assume a flat universe
with $\Omega_m = 0.27$, $\Omega_{\rm vac} = 0.73$ and $H_0 =
71$~km~s$^{-1}$~Mpc$^{-1}$, in which case $z=1$ corresponds to a
luminosity distance $D_L\approx6.6$~Gpc. We start our calculations
when afterglow phase starts, i.e. $m_0=M_{\rm ej}/\gamma_0$. This
happens at the deceleration radius $R_{\rm dec}=m_0/(\pi a_0^2 n
m_{\rm p})$, about $t_0 = R_{\rm dec} / (2 c \gamma_0^2)$ after the
main burst.

Figure 2 shows the effect of varying $R_{\rm rise}$ on the
lightcurves. A bump in the lightcurve appears when the jetted GRB
ejecta meets the boundary of stellar wind and the molecular cloud
(Region 2 in Figure 1). Following the bump is a marked decrease of
the brightness. The time the bump appears, $t_{\rm rise}$, is
strongly correlated with $R_{\rm rise}$. Therefore, from the time a
bump appeared on an observed GRB lightcurve, we can roughly estimate
the distance of the density boundary from the progenitor, provided
that other parameters are suitably chosen.

In Figure 3, we show the effect of density $n_2$ in Region (3) on
the lightcurves. We see that $n_2$ affects the shape of the bump in
the afterglow lightcurves and the level of the late time flux. The
larger $n_2$ is, the higher is the bump observed and the shorter is
the rising time of the bump.

We vary $d$ in figure 4 to demonstrate the effect of the width of
Region (2) on the lightcurves. This time a higher bump and a shorter
bump's rising time is caused by a smaller $d$. However, the
parameter $d$ cannot affect the level of the late time flux of the
afterglow.

In Figure 5, the effect of our viewing angle $\Theta$ on the
observed afterglow lightcurve is illustrated. As noted by previous
authors (e.g. Cheng et al., 2001), a viewing angle larger than
$1/\gamma$ will drastically suppress the early observed flux.

\section{Comparisons of the Model with GRBs 970508, 000301C, and 030226}

\subsection{GRB~970508}

GRB~970508 was detected simultaneously on 1997 May 8 at UT 21:41:50
by the Gamma-ray Burst Monitor (Costa et al., 1997) and the X-ray
Wide Field Camera on BeppoSAX (Jager et al., 1997). Its host galaxy
is a starburst galaxy (Sokolov et al., 1999). A high density shell
around the GRB progenitor of this burst has been suggested by Piro
et al. (1999). They interpreted the dense region to be the ejecta of
a preburst supernova. GRB~970508 has a gamma-ray fluence of
$3.1\times10^{-6}$~erg~cm$^{-2}$, a redshift $z = 0.835$, and a
luminosity distance $D_L = 5.30$~Gpc, corresponding to an isotropic
gamma-ray energy of $E_{\gamma,\rm iso} = 5.68\times10^{51}$~erg.

The optical flux remained constant (a plateau) for about 1
    day from early observations. Then the $R_c$--band flux increases $\sim 1.3$
    mag in $\approx 1 $ day. After that it decayed as a power law in the
    following tens of days (a possible break may occur around $t=25$~days). This peculiar behavior is puzzling. Panaitescu et al. (1998) and Chang
    et al.
    (2002) proposed that delayed energy injection from the central source could account for the
    bump. Figure~6 shows the observed $R_c$-band lightcurve of GRB~970508 and
the predictions by the cylindrical jet model using the parameters
listed in Table~1. The agreement is very good for over 100 days.
Specifically, the early bump at $1-2$ day after the burst can be
satisfactorily reproduced by our model. Thus, the early bump may
also be the results of density enhancement in the vicinity of a very
massive star in the context of the cylindrical jet model.

\subsection{GRB~000301C}

GRB~000301C was first detected with both the RXTE All-Sky Monitor
and the IPN spacecrafts Ulysses and NEAR on 2000 March 01 at UT
09:51:37 (Smith et al., 2000). The burst has a gamma-ray fluence of
$4.1\times10^{-6}$~erg~cm$^{-2}$, a redshift $z = 2.033$ and a
luminosity distance $D_L = 16.1$~Gpc, corresponding to an isotropic
gamma-ray energy of $E_{\gamma,\rm iso} = 4.17\times10^{52}$~erg.
Its infrared/optical afterglow lightcurves show a bump around day~3
and another bump around day~7. The first bump has been interpreted
by Garnavich et al. (2000) as due to gravitational microlensing,
while other explanations such as density enhancement (Berger et al.
2000) or energy injections have been suggested.

A comparison of the optical data of this burst with our model is
shown in Figure 7 using the parameters in Table 1. Our model can
reproduce the first bump which appears at around $t\sim3$ days. This
result gives support to a previous study (Berger et al., 2000) that
the bump is caused by density enhancement. Here we have only used a
simple density profile and we cannot reproduce the two bumps in the
lightcurve simultaneously. The two bumps may result from density
fluctuations within the high density shell.

\subsection{GRB~030226}

The gamma-ray burst GRB~030226 was first detected by the High Energy
Transient Explorer 2 satellite on 2003 February 26 at UT 03:46:31.99
(Suzuki et al., 2003). The gamma-ray fluence was $5.7\times 10^{-6}
$~erg~cm$^{-2}$. The burst has a redshift of $z = 1.98$,
corresponding to a luminosity distance $D_L = 15.54$~Gpc and an
isotropic gamma-ray energy of $E_{\gamma,\rm iso} =
5.53\times10^{52}$~erg. Owing to the rapid localization of the
burst, Fox et al. (2003) detected an optical counterpart only 0.11
day after the burst. It faded as $t^{-1.2}$ for $\sim0.2$~day,
re-brightened during $0.2-0.5$~day, and finally declined as
$t^{-2.0}$, as interpreted by Dai \& Wu (2003). Figure 8 shows the
comparison between our model using the numerical values shown in
Table 1 and the optical data of this burst.

Although there is no clear bump seen in the lightcurve of this
burst, our model really fits very well all data points. We have used
the photometric data of Pandey et al. (2004), which do not suffer
from possible inhomogeneous photometric calibrations in the GCN
Circulars.

\subsection{Summary of Comparisons with Models}

From the above fits, we summarize in Table 1 the values of the
parameters used in our calculations together with the deceleration
radius $R_{\rm dec}$. From Table 1, we see that $R_{\rm rise}$
ranges from $0.4-22$~pc. These values are consistent with the radius
($\sim 0.1-10$~pc) of pressure-driven bubbles created by strong
stellar winds of very massive stars at the end of their life (Weaver
et al., 1977; Shull, 1980). The values of $n_1=1.6-100$~cm$^{-3}$
used in our models are consistent with those calculated by Shull
(1980). Note that similar values are inferred by broadband afterglow
modeling employing conical jet geometry (Panaitescu \& Kumar, 2001,
2002; Yost et al., 2003). The values we obtained for $n_2$, ranging
from $n_2\sim10^2-10^3$~cm$^{-3}$, are typical interclump densities
in galactic GMCs. The parameter $d$ is the width of the ramping
density region from the stellar wind to the molecular cloud. We know
little about its ``standard'' values because this region is highly
viable from different works. It is primarily because of different
assumptions of the ambient densities of the massive stars and
evolutionary trends during the final stages of the massive stars.
Specifically, the nice fit of GRB~970508 (and the values of the
parameters used in the fit) suggests that the bubble is created by
the massive star during its main-sequence stage (see Weaver et al.,
1977).

\section{Discussion}

Long-duration GRB progenitors are widely believed to be massive
stars because their associations with SNe and star forming regions
is clearer than anytime before. Since molecular clouds are the only
places where very massive stars are born and die, we believe that
GRB progenitors are surrounded by stellar wind and overdense regions
resulting from wind interactions with dense molecular clouds.
Ramirez-Ruiz et al. (2001) also considered the afterglows of GRBs
resulting from such environments. They assumed that the jet of GRB
has a conical geometry and they were able to explain the late-time
re-brightenings in some GRB afterglow lightcurves, such as the bumps
observed in afterglows of GRB~970228, GRB~980326, and GRB~000911.
However, their model cannot explain the re-brightenings occurred at
$t\lesssim10$ days after the bursts, as seen in the optical
lightcurves of GRB~970508 and GRB~000301C.

We will first discuss the homogeneity of Region (1). It is well
known that stellar winds blow throughout the entire life of very
massive stars. This wind will dramatically affect the GRB
environment. Many authors (e.g. Chevalier \& Li, 1999, 2000; Wu et
al., 2003) have discussed how a free wind environment (i.e. $n
\propto r^{-2}$) affects the lightcurves of GRB afterglows. However,
the free wind region terminates at a radius much smaller than
$10^{16}$ cm if the star was born in a cloud of $n_{\rm
H}\sim10^{5}$ cm$^{-3}$ (Weaver et al., 1977; Shull, 1980). The
simulations by Ramirez-Ruiz et al. (2001) further demonstrated that
a region with nearly constant density (or quasi-uniform density)
indeed exists between the free wind region and the high density
region. Recently, this idea has also been suggested by several other
authors (e.g. Chevalier et al., 2004). Berger et al. (2003b) shows
that a radio flare from GRB~020405 gives support to the presence of
a uniform density medium in the range $R\sim10^{15}-10^{17}$ cm.

In this paper, we show the optical afterglow lightcurve signatures
of a GRB ejecta meeting a low-to-high density boundary. Our modeled
lightcurves are consistent with the afterglows of GRB~970508,
GRB~000301C, and GRB~030226. We employ the idea that the boundary is
formed by interactions between the preburst strong wind with a dense
cloud (Ramirez-Ruiz et al., 2001). It should be noted that other
possibilities may give rise to such a boundary as well. For example,
interactions between fast and slow winds (Luo \& McCray, 1991;
Vikram \& Balick, 1998) or preburst supernova ejecta (Vietri \&
Stella, 1998) may also trigger overdense regions (or shells) in the
surroundings of GRB progenitors. Thus, we consider such density
boundaries very likely to exist along the path of GRB remnants (see
e.g. Dai \& Lu, 2002) and our results can also apply to these
environments which involve such kind of density boundaries.

Reverse shock may be an important factor in cases when density jump
exists. However, in our current study, the reverse shock will not
form and can be essentially omitted. The reason is as follows. It is
believed that the condition to form a relativistic shock is that the
density increases abruptly, and that at the same time the density
contrast ($n_2/n_1$) must be much larger than 21 (Dai \& Lu, 2002;
Dai \& Wu, 2003). If this condition is not satisfied, then the
reverse shock is at most Newtonian (In this case, the emission from
the Newtonian reverse shock is very week as compared with that from
the forward shock, and can essentially be omitted. The dynamics of
the flow will not be affected markedly also). In our calculations,
the overall density contrast ($n_2/n_1$) is generally less than 21
(see our fits of GRB 000301C and GRB 030226). Additionally, in our
model, the density does not increase abruptly, but slowly. That is,
the density rises from $n_1$ to $n_2$ in a distance $d$ which is
between 0.5 pc and 5 pc. This is a relatively large distance. Thus,
the condition for relativistic reverse shock to take effects will
not be satisfied in the situations we considered here.

We notice that the afterglow of GRB~030329 also shows an early bump
at $1-2$ days. This bump can possibly be interpreted in our
framework, although an interpretation using a two-component jet
model also seems satisfactory (Berger et al., 2003a).

As a final note, the movement of a massive star during its life may
complicate the density distribution arising from the stellar wind
and its interaction with the surroundings. A better understanding of
the wind environments around very massive stars during the end of
their lives is important for a complete description of GRBs and the
afterglows.

We would like to thank the anonymous referee for useful comments and
suggestions, and X. Y. Wang for valuable discussions and comments.
This work is supported by a RGC grant of Hong Kong SAR Government of
China under HKU7014/04P.

\section*{References}

Ando, M. et al., 2003a. GCN Circ. 1882.\\
Ando, M. et al., 2003b. GCN Circ. 1884.\\
Berger, E. et al., 2000. ApJ 545, 56.\\
Berger, E. et al., 2003a. Nature 426, 154. \\
Berger, E., Soderberg, A.M., Frail, D.A., \& Kulkarni, S.R., 2003b.
ApJ 587, L5.\\
Biretta, J.A., Sparks, W.B., \& Macchetto, F., 1999. ApJ 520, 621.\\
Bloom, J.S., 2003. In: Gamma-Ray Bursts in the Afterglow Era --
Third Workshop, ASP Conf. Series, Vol. 2003, Feroci, M. et al., astro-ph/0303478.\\
Bloom, J.S. et al., 1999. Nature 401, 453.\\
Bloom, J.S. et al., 2002. ApJ 572, L45.\\
Castro-Tirado, A.J. et al., 1998. Science 279, 1011.\\
Castor, J., McCray, R., \& Weaver, R., 1975. ApJ 200, L107.\\
Chang, H.Y., Lee, C.H., \& Yi, I., 2002. A\&A 381, L5.\\
Cheng, K.S., Huang, Y.F., \& Lu, T., 2001. MNRAS 325, 599.\\
Cheng, K.S., \& Lu, T., 2001. ChJAA 1, 1.\\
Chevalier, R.A., \& Li, Z.Y., 1999. ApJ 520, L29.\\
Chevalier, R.A., \& Li, Z.Y., 2000. ApJ 536, 195.\\
Chevalier, R.A., Li, Z.Y., \& Fransson, C., 2004. ApJ 606, 369.\\
Costa, E. et al., 1997. IAUC 6649.\\
Covino, S. et al., 2003. GCN Circ. 1909.\\
Dai, Z.G., Huang, Y.F., \& Lu, T., 1999. ApJ 520, 634.\\
Dai, Z.G., \& Lu, T., 2000. ApJ 537, 803.\\
Dai, Z.G., \& Lu, T., 2002. ApJ 565, L87.\\
Dai, Z.G., \& Wu, X.F., 2003. ApJ 591, L21.\\
Esin, A.A., \& Blandford, R., 2000. ApJ 534, L151.\\
Fatkhullin, T., Komarova, V., Sokolov, V., Cherepashchuk, A., \& Postnov, K., 2003. GCN Circ. 1925.\\
Fendt, C., \& Ouyed, R., 2004. ApJ 608, 378.\\
Frail, D.A. et al., 2001. ApJ 562, L55.\\
Fox, D.W., Chen, H.W., \& Price, P.A., 2003. GCN Circ. 1879.\\
Galama, T.J. et al., 1998. ApJ 497, L13.\\
Galama, T.J. et al., 2000. ApJ 536, 185.\\
Galama, T.J., \& Wijers, R.A.M.J., 2001. ApJ 549, L209.\\
Garnavich, P.M., Loeb, A., \& Stanek, K.Z., 2000. ApJ 544, L11.\\
Garnavich, P.M., von Braun, K., \& Stanek, K.Z., 2003a. GCN Circ. 1885.\\
Garnavich, P.M. et al., 2003b. ApJ 582, 924.\\
Greiner, J., Ries, C., Barwig, H., Fynbo, J., \& Klose, S., 2003.
GCN Circ. 1894.\\
Guarnieri A. et al., 2003. GCN Circ. 1892.\\
Hjorth, J. et al., 2003. Nature 423, 847.\\
Holland, S., \& Hjorth, J., 1999. A\&A 344, L67.\\
Holland, S. et al., 2001. A\&A 371, 52.\\
Huang, Y.F., \& Cheng, K.S., 2003. MNRAS 341, 263.\\
Huang, Y.F., Dai, Z.G., \& Lu, T., 1999a. Chin. Phy. Lett. 16, 775.\\
Huang, Y.F., Dai, Z.G., \& Lu, T., 1999b. MNRAS, 309, 513.\\
Huang, Y.F., Dai, Z.G., \& Lu, T., 2000a. MNRAS, 316, 943.\\
Huang, Y.F., Dai, Z.G., \& Lu, T., 2000b. A\&A 355, L43.\\
Huang, Y.F., Gou, L.J., Dai, Z.G., \& Lu, T., 2000c. ApJ 543, 90.\\
Huang, Y.F., Tan, C.Y., Dai, Z.G., \& Lu, T., 2002. ChA\&A 26, 414.\\
Jager, R. et al., 1997. A\&AS 125, 557.\\
Krasnopolsky, R., Li, Z.Y., \& Blandford, R.D., 2003. ApJ 595, 631.\\
Lamb, D.Q., \& Reichart, D.E., 2000. ApJ 536, 1.\\
Luo, D., \& McCray, R., 1991. ApJ 379, 659.\\
Ma, H.T., Huang, Y.F., Dai, Z.G., \& Lu, T., 2003. ChJAA 3, 225.\\
Maiorano, E. et al., 2003. GCN Circ. 1933.\\
Masetti, N. et al., 2000. A\&A 359, L23.\\
Nysewander, M. C., Moran, J., Reichart, D., Henden, A., \& Schwartz,
M., 2003. GCN Circ. 1921.\\
Paczy\'{n}ski, B., 1998. ApJ 494, L45.\\
Panaitescu, A., \& Kumar, P., 2001. ApJ 560, L49.\\
Panaitescu, A., \& Kumar, P., 2002. ApJ 571, 779.\\
Panaitescu, A., M\'{e}sz\'{a}ros, P., \& Rees, M.J., 1998. ApJ 503,
314.\\
Pandey, S.B. et al., 2004. A\&A 417, 919.\\
Perdersen, H. et al., 1998. ApJ 496, 311.\\
Perley, R.A., Bridle, A.H., \& Willis, A.G., 1984. ApJS 54, 291.\\
Piran, T., 1999. Phys Rep. 314, 575.\\
Piro, L. et al., 1999. ApJ 514, L73.\\
Price, P.A., \& Warren, B.E., 2003. GCN Circ. 1890.\\
Ramirez-Ruiz, E., Dray, L.M., Madau, P., \& Tout, C.A., 2001. MNRAS 327, 829.\\
Ray, T.P. et al., 1996. ApJ 468, L103.\\
Reichart, D.E., \& Price, P.A., 2002. ApJ 565, 174.\\
Reichart, D.E., \& Yost, S.A., 2001. ApJ submitted, astro-ph/0107545.\\
Rhoads, J.E., 1997. ApJ 487, L1.\\
Rhoads, J.E., 1999. ApJ 525, 737.\\
Rhoads, J.E., \& Frutchter, A.S., 2001. ApJ 546, 117.\\
Rumyantsev, V., Biryukov, V., \& Pozanenko, A., 2003a. GCN Circ. 1908.\\
Rumyantsev, V., Biryukov, V., \& Pozanenko, A., 2003b. GCN Circ. 1929.\\
Rybicki, G.B., \& Lightman, A.P., 1979. Radiative Processes in Astrophysics, New York: Willey.\\
Sagar, R., Mohan, V., Pandey, S.B., Pandey, A.K., Stalin, C.S.,
\& Castro-Tirado, A.J., 2000. BASI 28, 499, astro-ph/0004223.\\
Sari, R., Piran, T., \& Halpern, J.P., 1999. ApJ 519, L17.\\
Schlegel, D.J., Finkbeiner, D.P., \& Davis, M., 1998. ApJ 500, 525.\\
Semkov, E.H., 2003. GCN Circ. 1935.\\
Shu, F.H. et al., 1995. ApJ 455, L155.\\
Shull, J.M., 1980. ApJ 238, 860.\\
Sokolov, V.V., 1999. A\&A 344, 43.\\
Solomon, P.M., Rivolo, A.R., Barrett, J., \& Yahil, A., 1987. ApJ
319, 730.\\
Suzuki, M. et al., 2003. GCN Circ. 1888.\\
Vietri, M., \& Stella, L., 1998. ApJ 507, L45.\\
Vikram, D., \& Balick, B., 1998. ApJ 497, 267.\\
Vlahakis, N., \& K\"{o}nigl, A., 2003a. ApJ 596, 1080.\\
Vlahakis, N., \& K\"{o}nigl, A., 2003b. ApJ 596, 1104.\\
von Braun, K., Garnavich, P., \& Stanek, K., 2003. GCN Circ. 1881.\\
Wang, X.Y., Dai, Z.G., \& Lu, T., 2000. MNRAS 317, 170.\\
Weaver, R., McCray, R., Castor, J., Shapiro, P., \& Moore, R., 1977.
ApJ 218, 377.\\
Woosley, S.E., 1993. ApJ 405, 273.\\
Wu, X.F., Dai, Z.G., Huang, Y.F., \& Lu, T., 2003. MNRAS 342, 1131.\\
Yost, S.A., Harrison, F.A., Sari, R., \& Frail, D.A., 2003. ApJ 597, 459.\\

\vspace{5mm}

\setlength{\tabcolsep}{0.09cm}
\begin{center}
\begin{table*}[htp]

\begin{tabular}
{|lcccccccccc|} \hline & $R_{\rm rise}$ & $n_1$ & $n_2$ & $d$ &
$M_{\rm ej}$ & & & & $\Theta$ & $R_{\rm dec}$
\\ GRB & (pc) & (cm$^{-3}$) & (cm$^{-3}$) & (pc) & ($10^{-8}M_{\odot}$) &
$p$ & $\xi_B^2$ & $\xi_e$ & (rad) & ($10^{17}$cm)
\\ \hline
\hline 970508 & 22 & 1.6 & 100 & 5 & 26 & 2.0 & $10^{-4}$ & 0.1 & 0.006 & $2.1\times10^{2}$\\
\hline 000301C & 2.95 & 10 & 60 & 1.5 & 0.19 & 2.5 & $10^{-3}$ & 0.01 & 0.01 & 0.23\\
\hline 030226 & 0.4 & 100 & 1200 & 0.5 & 0.80 & 2.1 & $1.5\times10^{-5}$ & 0.03 & 0 & 0.10\\
\hline
\end{tabular}\\ \\
Table 1: The parameters used in our model fits to optical afterglow
lightcurves of GRBs 970508, 000301C, and 030226
\end{table*}
\end{center}

\begin{center}
\begin{figure}[htb]
\includegraphics[height=3in]{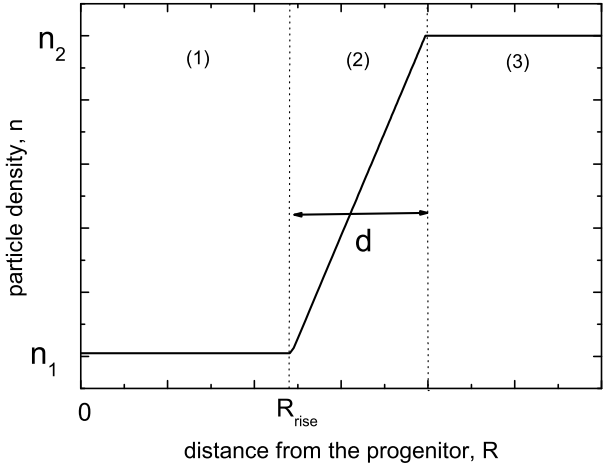}
\caption{The assumed density profile of the GRB system inside a
molecular cloud. The density in Region (1) is a constant $n_{1}$
while the density in Region (3) is $n_{2}$. Region (2) is an
intervening region at a distance $R_{\rm rise}$ from the progenitor
of width $d$.}
\end{figure}
\end{center}

\begin{center}
\begin{figure}[htb]
\includegraphics[height=3in]{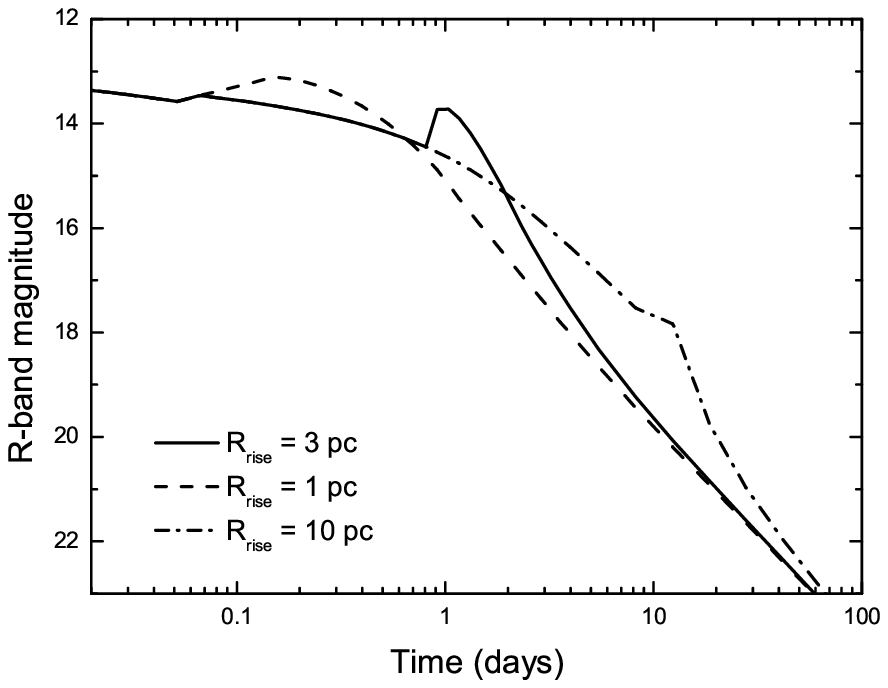}
\caption{Simulated optical lightcurves from our model. Except for
$R_{\rm rise}$, other parameters are kept constant (see text).}
\end{figure}
\end{center}

\begin{center}
\begin{figure}[htb]
\includegraphics[height=3in]{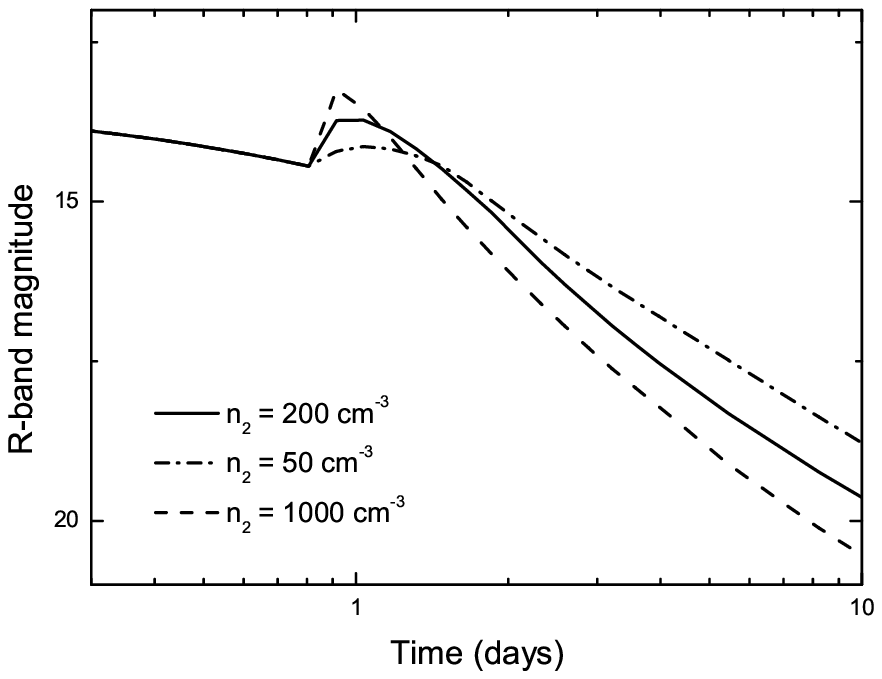}
\caption{Simulated optical lightcurves from our model. Except for
$n_2$, other parameters are kept constant (see text).}
\end{figure}
\end{center}

\begin{center}
\begin{figure}[htb]
\includegraphics[height=3in]{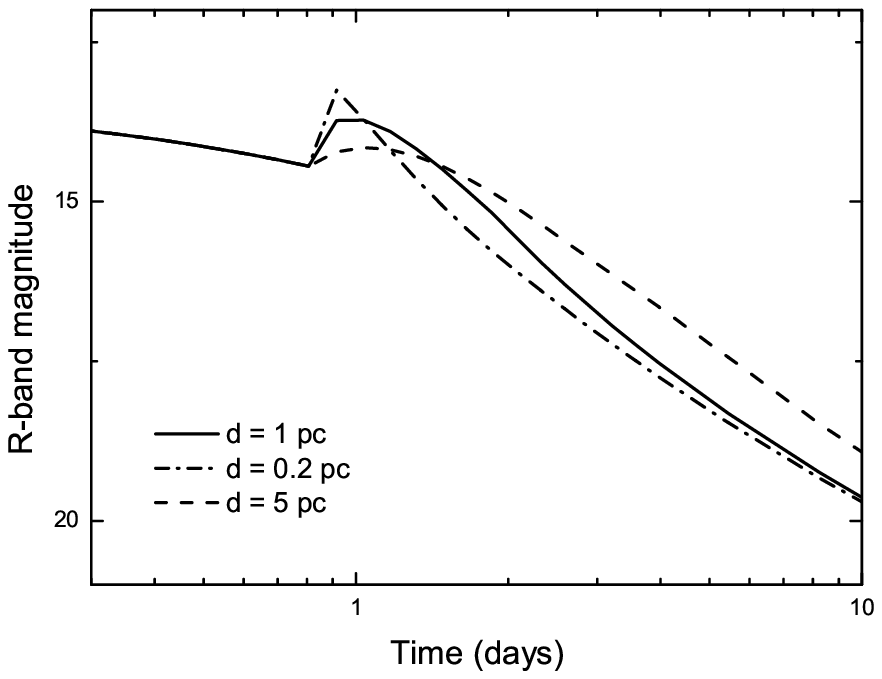}
\caption{Simulated optical lightcurves from our model. Except for
$d$, other parameters are kept constant (see text).}
\end{figure}
\end{center}

\begin{center}
\begin{figure}[htb]
\includegraphics[height=3in]{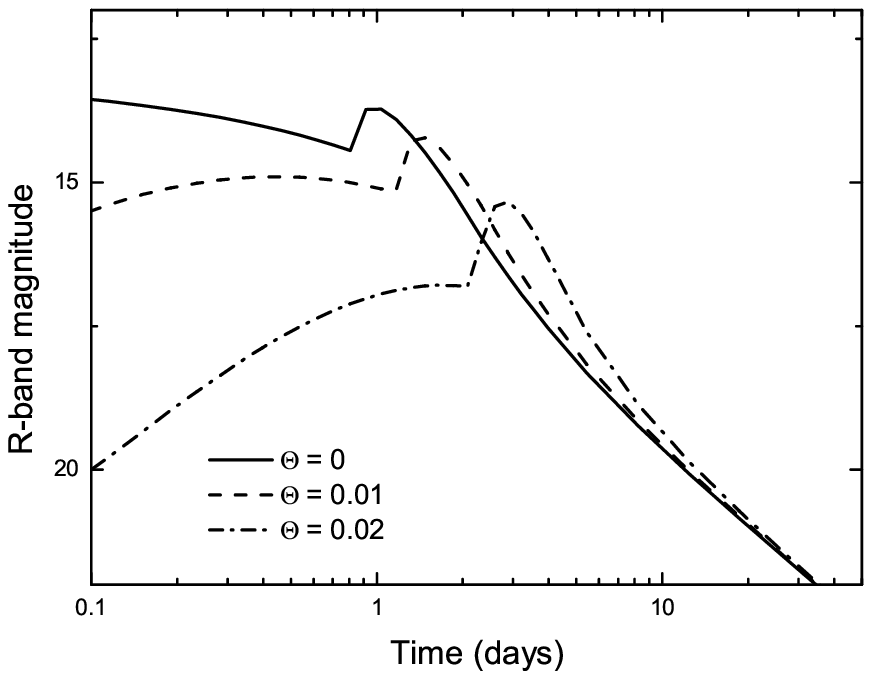}
\caption{Simulated optical lightcurves from our model. Except for
$\Theta$, other parameters are kept constant (see text).}
\end{figure}
\end{center}

\begin{center}
\begin{figure}[htb]
\includegraphics[height=3in]{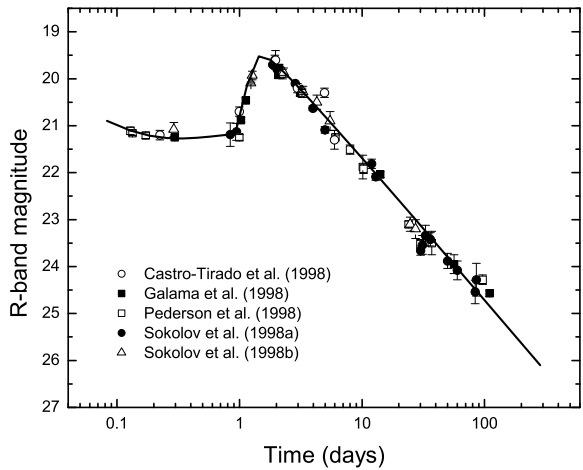}
\caption{Comparison of the model with GRB~970508 $R_c$-band
afterglow lightcurve. Data points are from Castro-Tirado et al.
(1998, empty circles), Galama et al. (1998, filled squares),
Perdersen et al. (1998, empty squares) and Sokolov et al. (1998a,
filled circles; 1998b, empty triangles). The error bar of the latest
data point is too small to be shown in the figure.}
\end{figure}
\end{center}

\begin{center}
\begin{figure}[htb]
\includegraphics[height=3in]{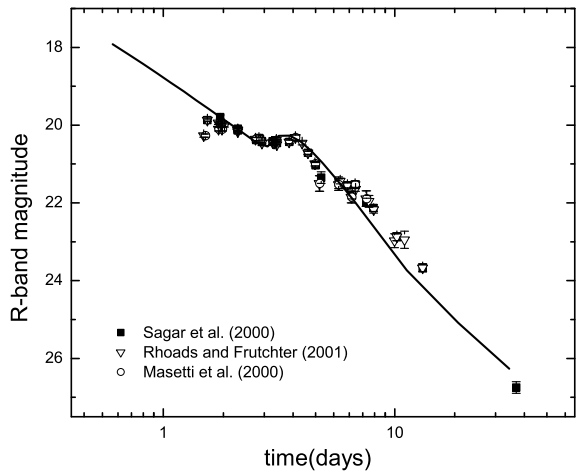}
\caption{Comparison of the model with GRB~000301C $R_c$-band
afterglow lightcurve. Data points are from Masetti et al. (2000,
empty circles), Rhoads \& Frutchter (2001, empty upper triangles)
and Sagar et al. (2000, filled squares), corrected for Galactic
foreground extinction ($R$ + 15\%; Schlegel et al., 1998).}
\end{figure}
\end{center}

\begin{center}
\begin{figure}[htb]
\includegraphics[height=3in]{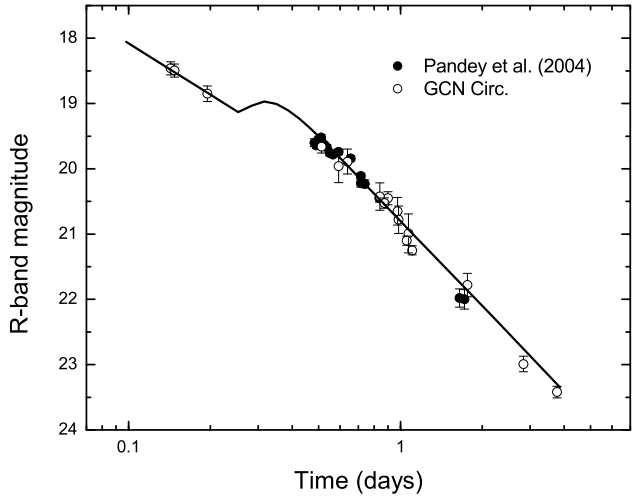}
\caption{Comparison of the model with GRB~030226 $R_c$-band
afterglow lightcurve. Data points are from Pandey et al. (2004;
filled circles) and the following GCN Circ. (empty circles): Ando et
al. (2003a, 2003b), Covino et al. (2003), Fatkhullin et al. (2003),
Garnavich et al. (2003a), Greiner et al. (2003), Guarnieri et al.
(2003), Maiorano et al. (2003), Nysewander et al. (2003), Price \&
Warren (2003), Rumyantsev et al. (2003a, 2003b), Semkov (2003) and
von Braun et al. (2003). }
\end{figure}
\end{center}

\end {document}